\documentclass{aa}
\input{epsf}
\begin{document}

   \thesaurus{08     
              (08.02.3;  
               08.06.1;  
               08.14.1;  
               08.09.2 XTE J1806--246;  
               13.25.5;  
               19.63.1)} 
   \title{Timing analysis of the X-ray transient source XTE J1806--246 (2S1803--245)}

   \author{M. Revnivtsev\inst{1,2}, K. Borozdin\inst{3,1},
   A. Emelyanov\inst{1,2}
          }

   \offprints{M.Revnivtsev (revnivtsev@hea.iki.rssi.ru)}

   \institute{Space Research Institute, Russian Academy of Sciences,
              Profsoyuznaya 84/32, 117810 Moscow, Russia
	\and
		Max-Planck-Institute f\"ur Astrophysik,
              Karl-Schwarzschild-Str. 1, 85740 Garching bei Munchen,
              Germany
	\and
		Los Alamos National Laboratory, NIS-2, Los Alamos, 87545 New Mexico, USA  
             }

   \date{}

   \maketitle

   \begin{abstract}

	An outburst of the X-ray transient source XTE J1806--246 (2S1803--245)
	was observed by the Rossi X-ray Timing Explorer from April-July
	1998. Strong quasi-periodical oscillations (QPO) with a central 
        peak frequency around 9 Hz was detected in
        one observation of the series performed by the PCA/RXTE experiment. 
	The X-ray flux from the source during the observation with QPO
	was maximal. 
        The energy spectrum of XTE J1806--246 at this time was
        softer than for other observations.  Statistically significant 
        variability of the QPO parameters was detected, in short-term
        correlation with the flux variability.  Fractional amplitudes of 
        Very Low frequency Noise and the QPO component of the power
	density spectrum  
        demonstrate a strong energy dependence, while other parameters 
        do not change significantly.

   \end{abstract}

\markboth{M.~Revnivtsev et al.: QPO in X-ray transient XTE J1806-246}{}
%

\section{Introduction}

The X-ray transient source XTE J1806--246 (=2S1803--245=MX1803--24) was
discovered by SAS--3 in May 1976 (\cite{jern_iau}, \cite{jern_nat}). 
An isolated X-ray burst from this region of the sky was detected by the
Wide Field Camera 1 (WFC1) on the BeppoSAX observatory on Apr. 2, 1998.
The All Sky Monitor (ASM) of the Rossi X-ray Timing Explorer satellite detected
the beginning of an X-ray outburst of the source on Apr. 16, 1998 
(\cite{asm_iauc}). Observations in other wavebands revealed 
the probable radio (\cite{radio}) and
optical (\cite{optics}) counterparts of the object.

Quasi Periodical Oscillations (QPO) were discovered by \cite{rudi}
in the power density spectrum (PDS) of the source in data obtained 
on May 3 1998. 
In this Letter we present results of timing analysis of the PCA/RXTE
experiment data, discuss the power density spectrum of the source 
and report detected correlations of QPO parameters on energy range 
and X-ray flux.

\section{Observations and analysis}

We analyzed archival data obtained by the RXTE observatory during 
the outburst of the source in Apr.-July 1998. Brief information 
about the observations is presented in the Table \ref{obslog}.
\begin{table}
\small
\caption{RXTE observations of XTE J1806--246
\label{obslog}}  
\begin{tabular}{ccccc}
\hline
\#&Obs.ID&Date, UT&Time start&PCA Exp.\\
&30412-01-..& & &sec\\
\hline
1&01-00&27/04/98&19:31:12&4832\\
2&02-..$^a$&28/04/98&18:59:12&1666\\
3&03-00&29/04/98&20:42:56&3253\\
4&04-00&03/05/98&21:01:20&2677\\
5&05-01&17/04/98&00:29:20&1371\\
6&05-00&17/05/98&12:42:24&3153\\
7&06-00&22/05/98&11:06:24&3255\\
8&07-00&08/06/98&07:59:28&2720\\
9&08-00&14/06/98&20:54:56&3403\\
10&09-00&01/07/98&19:45:04&991\\
11&10-00&07/07/98&19:50:08&1581\\
\hline
\end{tabular}

$^a$ -- This observation consists of several smaller ones: ..-02-00S,..-02-01S,..-02-02S,..-02-03S,..-02-04S
\end{table}

The data were analyzed according to the RXTE Cook Book recipes using the
FTOOLS, version 4.2 tasks. For background estimations we used the VLE
model for observations when X-ray flux was high, and the L7/240 for 
the observations corresponding to low flux.

The data were collected in different modes ({\em Standard 2}, {\em Event
Mode}, {\em Single Binned} and {\em Binned Mode}) with the best timing
resolution of 16$\mu$s in hard energy bands ($\ga$13 keV), and of
8 ms and 125 $\mu$s in soft energy bands ($\la$13 keV). The 
X-ray flux from the source appears to be highly erratic, with significant
variability at all time scales.

\begin{figure}
\epsfxsize=8cm
\epsffile[70 170 510 710]{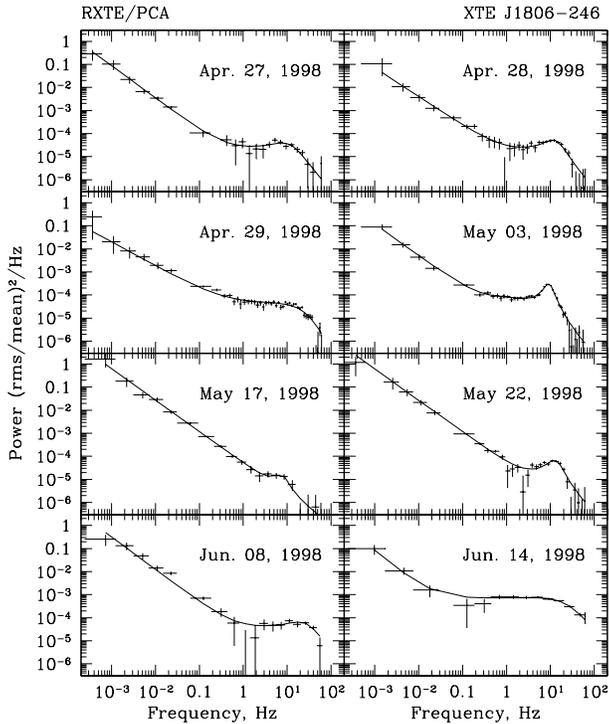}
\caption{The power density spectra of XTE J1806--246 during the different RXTE
observations (2--13 keV energy band). The solid lines represent models
for the PDS, that consist of Very Low Frequency Noise, High Frequency
Noise and QPO for observation \#4 (see text) 
\label{pds}} 
\end{figure}

To obtain a broad band power density spectra we used data of two
types. For the PDS at frequencies higher than 0.05 Hz we used 
16 sec data segments with an 8 ms time resolution. 
This gave us a set of power density spectra between 1/16 and
64 Hz, which were subsequently averaged and corrected for deadtime
modified poissonian noise
(\cite{deadtime}). To obtain PDS in a lower frequency band 
($\sim 5\times10^{-4}$ Hz -- $\sim$0.1 Hz) we used the 16-second time
resolution data of {\em Standard 2} mode, because it allowed us to take into
account the influence of the background variation on the PDS, which
might be of importance at such frequencies.  
 
\section{Results}

\subsection{Power density spectra}

\begin{figure}
\epsfxsize=8cm
\epsffile[20 140 560 710]{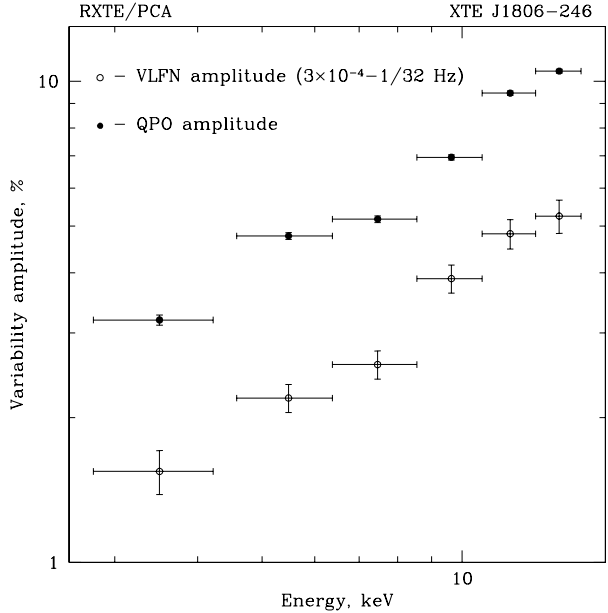}
\caption{Variability amplitudes for VLFN and QPO as functions of
energy (observation \#4).
\label{energy_dep}} 
\end{figure}

The power density spectra for different observations are presented in 
Fig.\ref{pds}.  The power law representing the Very Low
Frequency Noise (VLFN) component dominates the frequency band 
$10^{-4}$ -- 1 Hz for all PDSs.  The high Frequency Noise component
for frequencies higher than 1 Hz can be approximated either by another 
power law with an exponential cut-off at frequencies of 10--20 Hz or by
a wide Lorentzian. 
The slope of the VLFN component was in the range $\alpha\sim-1.0-1.5$ 
and its fractional variability was from 1.5 to 4 percent in the
energy band 2--13 keV. The amplitude of the VLFN fractional variability
increases with energy (see Fig. \ref{energy_dep} for observation \#4)
while the slope of the power law remains constant.

In the observation of May 3, 1998, a significant ($>10\sigma$) QPO peak 
was detected, agreeing with an earlier report by \cite{rudi}. 
The central frequency of the QPO, which was obtained by fitting 
a Lorentzian profile to the
power density spectrum averaged over the whole observation (2--13 keV
energy band) is equal to
$f=9.11\pm0.07$ Hz, with the width of the Lorentzian profile  $FWHM=5.4\pm0.2$
Hz. Its fractional variability amplitude in the frequency range
$\sim10^{-4}-64$ Hz was $4.7\pm0.1\%$. The amplitude of the QPO varies
with energy similar to the variation of the VLFN (Fig.\ref{energy_dep}). 

\begin{figure}
\epsfxsize=8cm
\epsffile[20 140 560 710]{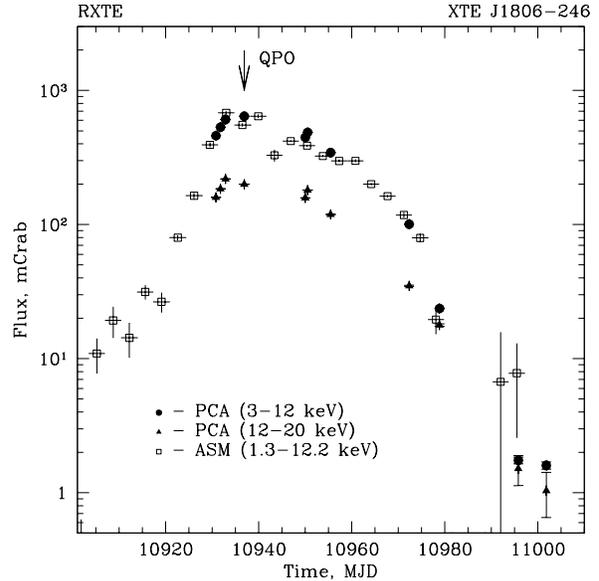}
\caption{The light curve of X-ray transient XTE J1806-246 during its 
outburst of 1998. Empty rectangles represent ASM data, filled 
points and triangles - PCA data in two energy bands. The date of
QPO detection is marked by the arrow.\label{lc}} 
\end{figure}

It is remarkable that this observation was taken
when X-ray flux from the source was the highest among all PCA observations.
The X-ray light curve of XTE J1806-246 during its outburst of 1998
is presented in Fig.\ref{lc}. It is worth noting that the difference 
in the X-ray flux detected in observations \#3 and \#4 is only $\sim$5\%,
but PDS are qualitatively different. 

\subsection{QPO parameters}

\begin{figure}
\epsfxsize=7.5cm
\epsffile[70 170 510 710]{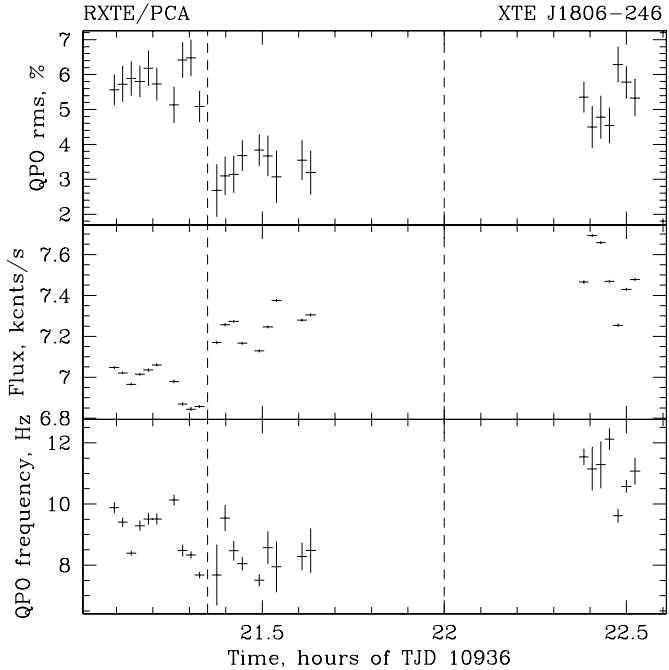}
\caption{The time dependence of the QPO parameters (frequency and
amplitude) and total X-ray flux variability during the observation
\#4. Dashed lines mark time intervals with different QPO/flux levels
(see text). 
\label{qpo}}
\end{figure}

In the observation of May 3, 1998 we detected significant changes
in the QPO parameters: the central frequency, width and amplitude 
of the fractional variability.  Fig. \ref{qpo} shows the variation of  
QPO amplitude and frequency in comparison with 
X-ray flux variation (all values were computed for the energy band 3--13 keV). 

\begin{figure}
\epsfxsize=7.5cm
\epsffile[55 360 500 710]{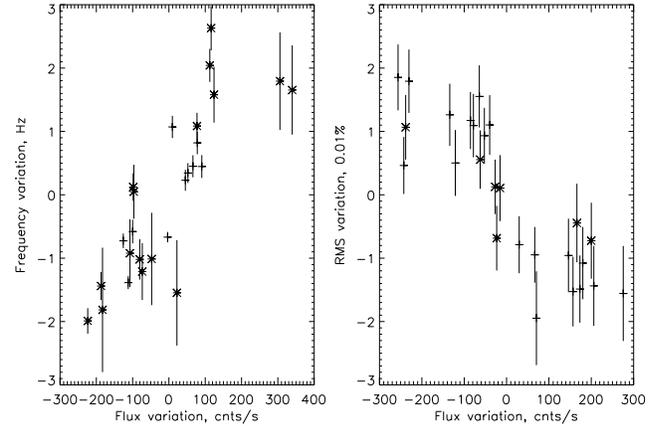}
\caption{The correlations between X-ray flux and QPO parameters for
short time intervals. On the left panel data for the second and third
intervals are marked by stars. On the right panel data for the third
interval are marked by stars. 
\label{corr}}
\end{figure}

Because both the X-ray flux from the source and QPO parameters have
demonstrated time variability, we tried to look for correlations 
between them, which might be important for the overall understanding of 
the QPO phenomenon.  One might note three distinctive time intervals 
during the observation of May 3. The first corresponds to 
``low'' flux (below 7100 cnts/s) and  ``strong''
QPO (rms higher than 5\%) with ``low'' frequency (below 10 Hz). 
In the second interval the flux is ``medium'' (7100 -- 7400 cnts/s),
QPO is ``weak'' (rms $\la 5\%$) and still of ``low'' frequency.
During the third interval the flux is ``high'' ($\ga$ 7400 cnts/s),
QPO is ``strong'' and of a higher frequency (10-12 Hz).  While there is
no global correlation between these three intervals, we have found 
some evidence for short-term correlations within each of them.
The results are presented in Fig.\ref{corr}.  Parameter variations were 
calculated by subtraction of the mean value for each interval. 
For flux-frequency dependence data for the first interval were separated from
later data.  In order to obtain flux-rms dependence the first two
intervals were analyzed together, and the third one - separately.  
A flux-frequency correlation and  flux-rms anticorrelation are clearly
seen at short scales, but an influence of other, not related to QPO, 
processes might be the reason why such correlations could not be 
expanded for longer time intervals.  We would like to note that 
the measurement of total X-ray flux and the determination of QPO parameters
was done by completely independent methods, so any correlation between them
must be not methodical, but physical.
We were concerned about a possible systematical cross-correlation 
between QPO amplitude and QPO frequency values, both determined 
in the same procedure, but our analysis has not revealed any 
evidence of such a correlation.

\subsection{Color-color diagram}
 
\begin{figure}
\epsfxsize=8cm
\epsffile[10 160 570 710]{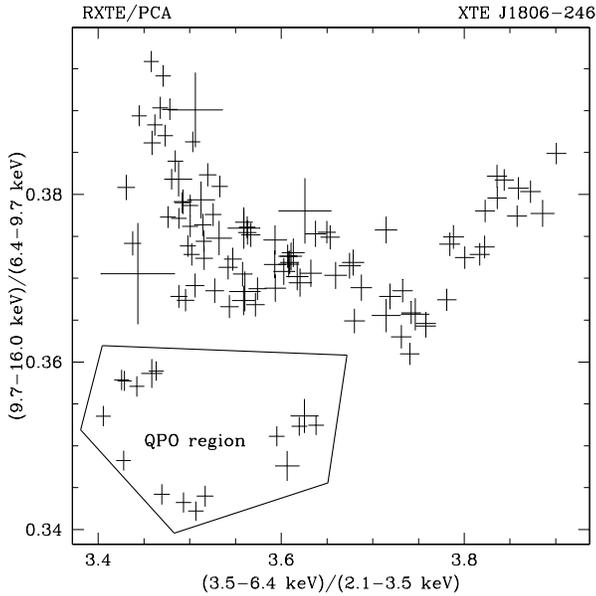}
\caption{The color-color diagram of XTE J1806-246 during the first 8
    observations. The points within 
    the solid line polygon correspond to the time when the strong QPO was
    visible on the PDS spectra\label{ccd}}
    \end{figure}

We used data in different energy channels to build a 
color-color diagram (CCD), which allows one to follow
spectral changes of the source (\cite{zoo}).  
The CCD for the data of the first 7 analyzed observations 
is shown in Fig.\ref{ccd}.
For this CCD we used four energy bands in the PCA spectrum: 2.1--3.5 keV,
3.5--6.4 keV, 6.4--9.7 keV and 9.7--16.0 keV.

The overall shape of the points distribution reminds us of the diagrams for 
some Z-sources, such as GX17+2 and Sco X-1 (see \cite{zoo}).  
However, we did not find a clear correlation between QPO and 
one of the branches on diagram.  Instead, we found that the QPO 
region corresponds to a lower value of hard color regardless of 
the value of soft color.

\section{Discussion}

Quasi-periodical oscillations in power density spectra have been detected
from many sources of different natures (see \cite{klis95} for review and 
references therein).  We discuss here a strong QPO in the PDS of 
the X-ray source XTE J1806-246, which can be considered a typical
X-ray transient.  The nature of this source is 
more or less clear because of the definite detection of the X-ray burst from
it (\cite{sax_iau}), so one can consider XTE J1806-246 a neutron
star with a low magnetic field. X-ray bursts are more typical for atoll
sources (see the classification in \cite{zoo}). 
However, the shape of the color-color diagram, PDS dominated by VLFN and 
HFN components and the fact that QPO was observed at the maximum X-ray
light curve of the source is strongly suggestive for  
the classification of XTE J1806-246 as another Z-source, known 
to be neutron stars, probably with a magnetic field stronger than that
of typical X-ray bursters, but weaker than that of pulsars (\cite{klis94}).
As mentioned by \cite{rudi} it would be then the first ever known
Z-transient.  Other reputed Z-sources are all persistent X-ray sources
(\cite{zoo}).

The origin of QPOs in Z-sources remains controversial. Some of models 
proposed include the beat frequency model (\cite{alpar}), hot spots 
in a boundary layer (\cite{hameury}), obscuration of the central X-ray 
source by an accretion disk (\cite{stella}) etc. (for review see 
\cite{lamb88}, \cite{klis95} and references therein).
The correlation between the X-ray flux and the QPO frequency 
we found for short time scales could be an indication 
that the frequency of QPO is linked with the typical radius
of the accretion disk in the system, which in turn correlated with 
the luminosity of the source. An anticorrelation between the
source flux and the amplitude of fractional 
variability associated with QPO demonstrates that short-term increases
of X-ray flux (microflares) might be caused by a mechanism, not
related with QPO. 

The Very Low Frequency Noise was detected in all observations of XTE J1806-246
by PCA/RXTE, in all energy bands with the increasing amplitude 
of fractional variability towards higher energies. 
The striking similarity in the energy dependences of
QPO and VLFN amplitudes suggests that these components originate in
the same region of the binary system.  The energy spectrum of the source
is probably formed by hard highly variable component and soft constant
component, which explains the increase of QPO and VLFN amplitudes 
with energy.

\begin{acknowledgements}

This research has made use of data obtained through the High Energy
Astrophysics Science Archive Research Center Online Service, provided 
by the NASA/Goddard Space Flight Center. The work has been supported 
in part by RFBR grant 96-15-96343. Authors would like to acknowledge
helpful comments of anonymous referee.
We are grateful to Ms.K.O'Shea for the language editoring of the manuscript.

\end{acknowledgements}

\end{document}